# A trial emulation approach for policy evaluations with group-level longitudinal data

Eli Ben-Michael[1], Avi Feller[1,2], and Elizabeth A. Stuart[3]

Forthcoming at *Epidemiology*

1 Department of Statistics, University of California, Berkeley
357 Evans Hall
Berkeley, CA 94720-3880
ebenmichael@berkeley.edu

2 Goldman School of Public Policy, University of California, Berkeley
2607 Hearst Avenue
Room 309
Berkeley, CA 94720
(510) 642-2067
afeller@berkeley.edu

3 Department of Mental Health, Johns Hopkins Bloomberg School of Public Health
624 N. Broadway
Room HH839
Baltimore, MD 21205 USA
410-502-6222
estuart@jhsph.edu

1) Eli Ben-Michael was supported by a National Science Foundation Research and Training Grant #1745640, 2) Dr. Feller's time was supported by the Institute of Education Sciences, U.S. Department of Education, through Grant R305D200010. 3) Dr. Stuart's time was supported by the National Institutes of Health through the RAND Center for Opioid Policy Tools and Information Center (OPTIC; P50DA046351).

Replication data and code are available at https://github.com/ebenmichael/policy-trial-emulation.

We thank Elizabeth Stone, Ian Schmid, and Elena Badillo Goicoechea at Johns Hopkins for editorial help and constructive comments.

**Abstract**

To limit the spread of the novel coronavirus, governments across the world implemented extraordinary physical distancing policies, such as stay-at-home orders, and numerous studies aim to estimate their effects. Many statistical and econometric methods, such as difference-in-differences, leverage repeated measurements and variation in timing to estimate policy effects, including in the COVID-19 context. While these methods are less common in epidemiology, epidemiologic researchers are well accustomed to handling similar complexities in studies of individual-level interventions. "Target trial emulation" emphasizes the need to carefully design a non-experimental study in terms of inclusion and exclusion criteria, covariates, exposure definition, and outcome measurement — and the timing of those variables. We argue that policy evaluations using group-level longitudinal ("panel") data need to take a similar careful approach to study design, which we refer to as "policy trial emulation." This is especially important when intervention timing varies across jurisdictions; the main idea is to construct target trials separately for each "treatment cohort" (states that implement the policy at the same time) and then aggregate. We present a stylized analysis of the impact of state-level stay-at-home orders on total coronavirus cases. We argue that estimates from panel methods — with the right data and careful modeling and diagnostics — can help add to our understanding of many policies, though doing so is often challenging.

## Introduction

To limit the spread of the novel coronavirus, governments across the world implemented extraordinary "non-pharmaceutical interventions," such as closing non-essential businesses and imposing quarantines. The specific policies — and the decisions to lift them — varied widely, with particularly dramatic variation within the United States.[1] Learning about the impact of these policies is important both to inform future policy decisions and to construct accurate forecasts of the pandemic.

There are well-established epidemiologic methods for estimating the impact of an intervention that occurs in a single location and at a single time point, dating back to John Snow and the cholera epidemic in London.[2][3] Often known as *difference-in-differences*, the basic approach constructs counterfactual outcomes using group-level longitudinal data from the pre- and post-policy change time periods and from localities (e.g., states) that did and didn't implement the policy. Variants of this approach are known as *panel data methods* and include *event studies* and *comparative interrupted time series* models.

There is no clear consensus in epidemiology, however, on how to proceed when many localities implement a policy over time, sometimes known as *staggered adoption*. Fortunately, epidemiologic researchers are well accustomed to handling similar complexities in studies of individual-level interventions, with decades of research on strong non-experimental study designs, including methods that handle confounding and variation in timing of treatment across individuals. "Target trial emulation" emphasizes the need to carefully design a non-experimental study in terms of inclusion and exclusion criteria, covariates, exposure definition, and outcome measurement —
and the timing of all of those variables.[4][5]

In this paper, we argue that policy evaluations using panel data need to take a similarly careful approach to study design, which we refer to as "policy trial emulation." The main idea is to construct target trials separately for each "treatment cohort" (states that implement the policy at the same time) and then aggregate.[6] We illustrate this approach by presenting a stylized analysis of the impact of state-level stay-at-home orders on total coronavirus cases. We believe this new connection to trial emulation is an important conceptual advance, though the underlying statistical methods we discuss are well-established,[7] and many of these points have been made in other contexts.[8] We argue that estimates from panel methods — with the right data and careful modeling and diagnostics — can help add to our understanding of policy impacts. The underlying assumptions, however, are often strong and the application to COVID-19 anti-contagion policies is particularly challenging.

**The elements of "policy trial emulation"**

We now describe key steps in conducting "policy trial emulation," which are necessary and obvious when designing a randomized trial and are becoming more common in the design of non-experimental studies. We argue that these steps are just as important when evaluating policies with aggregate longitudinal panel data.

We illustrate the key idea with a stylized policy evaluation: measuring the impact of US states adopting a "shelter in place" or "stay-at-home" policy on COVID-19 case counts. These orders urge or require citizens to remain at home except to conduct essential business, e.g., grocery shopping or exercise; we use NYT Tracker to define policy enactment dates and obtain daily case counts.[9] No IRB approval was needed for this use of publicly available aggregate data.

**Defining units and exposures**

First, we must have a consistent definition of the exposure. Specifically, there is only one form of treatment, and the outcome we see under a particular policy environment is equal to the potential outcome under that policy environment.[10] For stay-at-home policies, different states enacted different requirements that broadly fall under this header, and the NYT definition is just one. This introduces a trade-off. We could consider multiple types of treatment separately, such as closing schools versus closing non-essential businesses. However, this greatly expands the dimensionality of the specific exposure under consideration. Instead we consider "packaging treatments" together, allowing for some variation in the specific implementation of the policy across units. As a result, the estimated effect averages over different policy-specific effects within the data, which may be less interpretable and may violate the consistency assumption.[8,11]

Second, there is growing evidence that stay-at-home orders had only modest impacts on individual behavior — in many states, individuals reduced their mobility even in the absence of official policy changes.[12] Thus, we focus here on emulating an intent-to-treat (ITT) analysis, where individuals are randomized to treatment conditions, but the amount of treatment actually received (e.g., the dose) may differ across people. The ITT is often relevant for examining whether the policy is effective overall, regardless of specific implementation. If we had data on, e.g., state-wide implementation of the policy or the level of adherence as determined by mobility measures, we could conduct an analysis analogous to a per protocol effect, e.g., estimating the effect of "full" policy implementation (i.e., all individuals following the stay at home guidelines).[13] The trial emulation framework helps clarify the additional assumptions necessary for conducting a per protocol analysis in this context.

Finally, an important complication is reasoning about interactions between units, e.g. people with COVID-19 traveling across state lines and spreading infection. The trial emulation framework makes clear that we must pay close attention to this when defining our target trials: nearly all standard tools for policy evaluation assume no interference between units, that is, a state's outcome only depends on that state's intervention. Violations of this assumption, sometimes known as spillovers or contagion, complicate the definitions of units and exposures and can lead to bias.[14] Understanding and explicitly modeling these violations is paramount when studying policies to control infectious diseases. For example, Holtz et al. use travel and social network information to account for spillovers across states.[15] We refer interested readers to the large literature on observational causal inference with interference in general[16] and on panel data methods in particular.[17]

**Causal contrasts of interest**

After defining units and exposures, the next step in the trial emulation framework is to define the estimand of interest.[13] As discussed above, we focus on the ITT for treated states. Formally, let $W_{it}$ be an indicator that state *i* has a stay-at-home order at time t; and let $Y_{it}$ be the corresponding observed outcome. We can express the causal quantity of interest via potential outcomes: $Y_{it}$ ($W_{it} = 1$) is the outcome if the stay-at-home order is enacted, and $Y_{it}$ ($W_{it} = 0$) is the outcome if the order is not enacted. The causal contrast of interest is then a difference between these potential outcomes, $Y_{it}(1) - Y_{it}(0)$, averaged over the states that implemented the policy and over post-treatment time periods.[7]

We also focus on the impact of "turning on" these policies, but of course states also turn them "off." Just as in the individual exposure case, modeling individuals or locations turning exposures on AND off is complex. If that seems ambitious in a trial setting it is often even more ambitious in a non-experimental context!

**Defining time zero**

The next step in trial emulation is to define "time zero", i.e., the point in time when individuals or states would have been randomized to treatment conditions.[6] This is crucial for clearly distinguishing baseline (pre-treatment) measures from outcomes (post-treatment): inappropriately conditioning on or selecting on post-treatment variables can cause as much bias (including immortal time bias) as can confounding.[5 18].

In standard target trial emulation, time zero is often defined based on when individuals meet the specified eligibility criteria and is applied equally to both treated (exposed) and control (unexposed) units. In policy trial emulation, states are often "eligible" to implement a policy at any point, though this can also occur in standard target trial emulation.[13] For treated states, we typically use the date the policy is enacted as time zero; for comparison states, however, analysts essentially need to identify the moment in time that a state "could have" implemented the policy but did not.

One option is to align states based on calendar time. For instance, below we focus on a target trial for states that enact stay-at-home orders on March 23; we similarly set March 23 as "time zero" for the comparison states. In the COVID setting, however, we might instead want to measure time since the start of the pandemic in a specific location --- given the sudden emergence of the pandemic, case counts are essentially undefined before this time. This presentation is in line with many of the epidemiologic models.[19] We refer to this as *case time* and, as an illustration, index time by the number of days since the 10th confirmed case.

Figure 1 (left panel) shows the timing of statewide orders in calendar time beginning in mid-March with "early adopters" such as California, New Jersey, Illinois, and New York. Statewide orders continued through early April, with "late adopters" including Florida and Alabama. Several states, including Iowa and Arkansas, never enacted a state-wide order. Figure 1 (right panel) shows the timing in case time. From this perspective, early adopters, including West Virginia and Idaho, enacted stay-at-home orders within days of the tenth case, while California — the first state to enact a statewide stay-at-home order — was relatively late to do so.

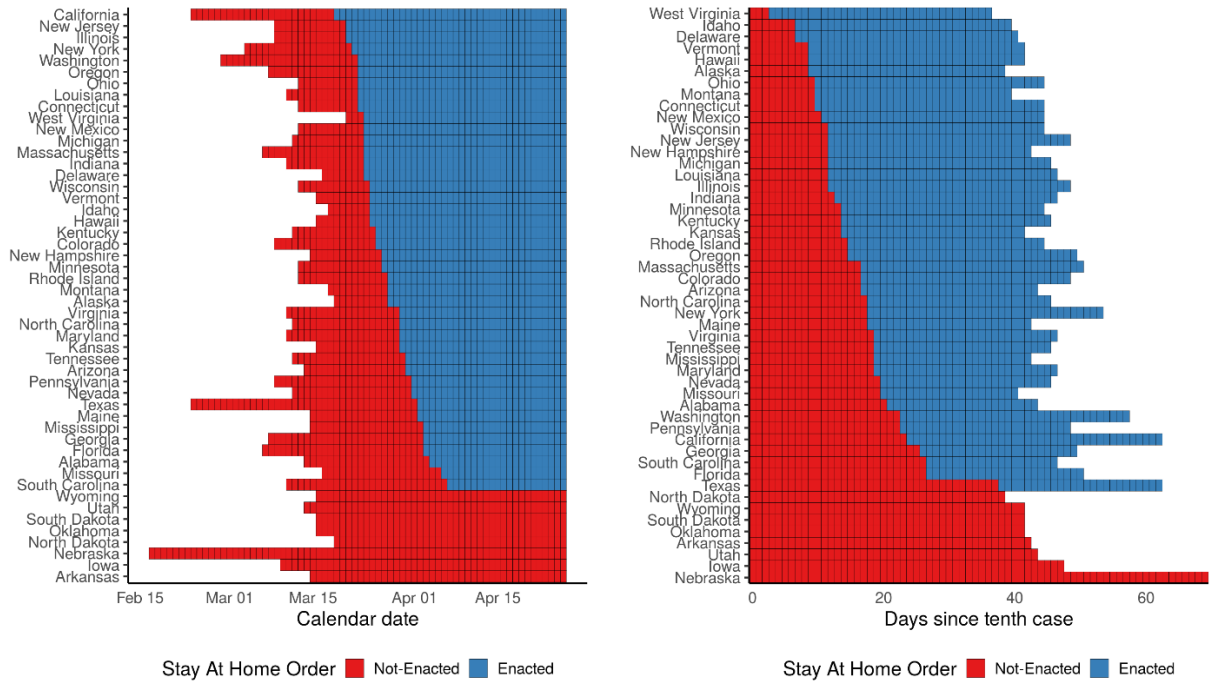

**Figure 1:** Timing of statewide stay-at-home orders, by calendar date (left) and case time (right). Calendar dates with fewer than 10 cases and case times after April 26th, 2020 are unshaded.

The choice of time zero will depend on the context. In the COVID setting, case time is more consistent with models of infectious disease growth but is also more sensitive to measurement error. Calendar time is more natural for accounting for "common shocks" to states, such as changes in Federal policy, which occur on a specific date. Moreover, for some specific questions, such as impacts on employment outcomes, this distinction might not be relevant. For ease of exposition, we focus on calendar time in the main text and give analogous results in case time in the Appendix.

**Defining outcomes**

The next step in policy trial emulation is to clearly define the outcomes, both the measures themselves and their timing ($t$ in the notation above). In a typical trial, the outcomes might be something like "mortality within 6 months." In our COVID case study we focus on two different outcome measures: (a) the (log) number of cases, and (b) the log-ratio of case counts from the previous day. The first is a measure of the cumulative effect, while the latter is a measure of the day-by-day changes in growth. We focus on log transformed data because exponential disease growth can result in different pre-intervention trends on the raw outcome scale; we further discuss the risk of pre-trends below, and present results for raw case counts and case growth in the Appendix. Data quality is also a key concern. Care needs to be taken to select outcomes that can be measured accurately; in particular, differential changes in the testing regime across states over time can lead to the illusion of an effect. Finally, we focus on outcomes that are measured at the aggregate (e.g., state) level, which is natural given the outcomes we examine. However, it is sometimes more reasonable to consider targeting a *cluster*- (rather than individual-) randomized trial in which treatment occurs at some aggregate level but outcome data are measured at the individual level data.[19] The same trial emulation principles apply, and hierarchical modeling can be used to account for the multilevel structure.

## Single target trial

We now describe a single target trial, and then turn to describing a sequence of nested target trials, which more fully captures the staggered adoption setting. Here the goal is to estimate the impact for a single *cohort* of states that enact a stay-at-home order at the same time. For now, we focus on the five states that enacted the policy on March 23, 2020: Connecticut, Louisiana, Ohio, Oregon, and Washington.

**Selecting treated and comparison states**

In a randomized trial, researchers have the "luxury" of knowing that the exposed and unexposed units are similar on all baseline characteristics in expectation. This is not the case in non-experimental studies such as policy evaluations. Thus, a key step in policy trial emulation is carefully selecting comparison units. We discuss statistical approaches for adjusting for differences in the next section.

An important consideration in selecting comparison units is the length of follow up — once a comparison state enacts the treatment, it is no longer a useful comparison without strong modeling assumptions. Only 19 days passed between when the first and last states enacted stay-at-home orders, and if we compare the March 23 cohort to late-adopting states, we can observe effects for at most 10 days. In general, across cohorts, the longer the follow up, the fewer available comparison states.[*]

Due to the lag between virus exposure, symptoms, and testing, we expect stay-at-home-orders to have delayed or gradual effects on the number of confirmed cases and case growth. How we define the start of the "outcome" time period (to then allow for different patterns of effects) is intimately related to the question of defining time zero, discussed above. Therefore, we compare the treated cohort to the eight "never treated" states, allowing for estimates of longer-term effects.[†] In principle, we could use "not yet treated" states in the comparison group, dropping states from the comparison group at the time they enact the policy. However, the set of "not yet treated" states will change throughout the follow up period. It may then be difficult to assess whether changes in effects are merely due to the changing composition of the comparison states. Additionally, for each set of comparison states at each follow up period we will need to perform the diagnostic checks we describe below, potentially leading to an unwieldy number of diagnostics.

---

[*] The choice of time scale is especially important if comparing with late adopters, since a state might be a valid comparison in calendar time but not case time: Ohio's stay-at-home order was enacted after California in calendar time but before California in case time.

[†] These are Arkansas, North Dakota, South Dakota, Iowa, Nebraska, Oklahoma, Wyoming, and Utah.

**Estimating treatment effects**

Once we have specified the target trial, the final stage is estimating the treatment effects and evaluating diagnostics for underlying assumptions. While similar to standard trial emulation contexts in many ways, there are particular nuances and complications in the policy trial emulation setting given the longitudinal time-series nature of the data and the relatively small number of units (e.g., 50 states). We illustrate this setup with simple estimators, especially the canonical "difference-in-differences" estimator; we discuss alternative estimators below.

**Table 1:** Average log growth rate in daily case counts for the March 23 Cohort and the never treated states (% day-over-day growth in parentheses). The pre-period is from March 8 to March 22; the post period is from March 23 to April 26.

**Stay-at-Home Order**

|  | **Pre** | **Post** | *Difference* |
|---|---|---|---|
| **March 23 Cohort** | 0.31 (37%) | 0.09 (10%) | *-0.22 (-20%)* |
| **Never Treated Cohort** | 0.24 (27%) | 0.10 (11%) | *-0.14 (-12%)* |
| *Difference* | *+0.07 (+10%)* | *-0.01 (-1%)* | ***-0.08 (-8%)*** |

**Difference-in-Differences Fundamentals.** The basic building block of traditional panel data estimation is "difference-in-differences" (DiD). To build up to the DiD estimator, consider two possible (flawed) comparisons. We could compare the growth rate in the March 23 cohort before and after the stay-at-home order; Table 1 shows these growth rates, implying a decrease in the average log growth rate of 0.22, a reduction of about 20 percentage points. However, this simple comparison relies on the heroic assumption that the stay-at-home order is the only change affecting the growth rate following March 23. Instead, we could directly compare the post March 23 growth rates for the two cohorts. From March 23 to April 26 the treated states' average log growth rate was 0.01 (1 percentage point) lower than the never treated states. While this approach protects against shared "shocks" between the two cohorts — e.g., changes in national policy or testing — it does not adjust for any differences in pre-intervention cases.

Difference-in-differences combines these two approaches. First take the pre/post estimate of a decrease in the log growth rate by 0.22 in the March 23 cohort and compare it to a pre/post change in the never treated states, a decrease of 0.14. Taking the difference of these differences (hence "difference-in-differences") yields an estimated reduction of the log growth rate by 0.08, or an 8-percentage point decrease in the growth rate (Table 1). Formally, this estimator is:

$$\widehat{DID}_g = \left(\overline{Y_{1g}} - \overline{Y_{0g}}\right) - \left(\overline{Y_{1\infty}} - \overline{Y_{0\infty}}\right)$$

where $\overline{Y_{0g}}$ and $\overline{Y_{1g}}$ denote the pre- and post-treatment average outcomes for cohort $g$, and $g = \infty$ denotes the never treated cohort.

**Assumptions.** The key to the differences-in-differences framework is a *parallel counterfactual trends* assumption: loosely, in the absence of any treatment, the trends for the treated cohort would be the same as the trends for the never-treated states, on average.[3][20] This assumption is inherently dependent on the outcome scale: if trends in *log* cases are equal, then trends in "raw" case numbers cannot also be equal. This assumption would be violated if:

- **Anticipation**: behavior changes *before* the state-wide shutdown, since pre-time zero measures would no longer truly be "pre-treatment." In the case of stay-at-home orders, there is strong evidence of such anticipatory behavior.[12]

- **Time-varying confounding**: the policy implementation decision making process depends on features other than baseline levels, i.e., the average level of the outcome of interest in the baseline time period. For example, this would be violated if governors enacted stay-at-home orders in response to *trends* in case counts.

Finally, this approach relies entirely on outcome modeling and therefore differs from many common methods in epidemiology that instead model treatment, especially inverse probability of treatment weighting (IPTW). Recent "doubly robust" implementations of difference-in-differences also incorporate IPTW and therefore rest on different assumptions than outcome modeling alone, including the positivity assumption that all units have a non-zero probability in being in each of the treatment conditions. Standard difference-in-difference models avoid positivity by instead relying on a parametric outcome model that potentially extrapolates across groups.[7][22]

**Diagnostics and allowing effects to vary over time.** The basic 2x2 table DiD estimator is a blunt tool: it estimates the effect averaged over the entire post-treatment period. By combining various 2x2 DiD estimators we can estimate how effects phase in after March 23. First, we pick a common reference date, often the time period immediately preceding treatment (here, March 22). Then for every other time period we estimate the 2x2 DiD relative to that date. Concretely, to estimate the effect $k$ periods before/after treatment we compute the 2x2 estimator:

$$\widehat{DID}_{kg} = \left(\overline{Y_{kg}} - \overline{Y_{-1g}}\right) - \left(\overline{Y_{k\infty}} - \overline{Y_{-1\infty}}\right),$$

where $\overline{Y_{kg}}$ is the average for cohort $g$, $k$ periods before/after treatment. Figure 2 shows these estimates for the March 23 cohort, sometimes known as *event study plots*,[23] with uncertainty quantified via a leave-one-unit-out jackknife.[7]

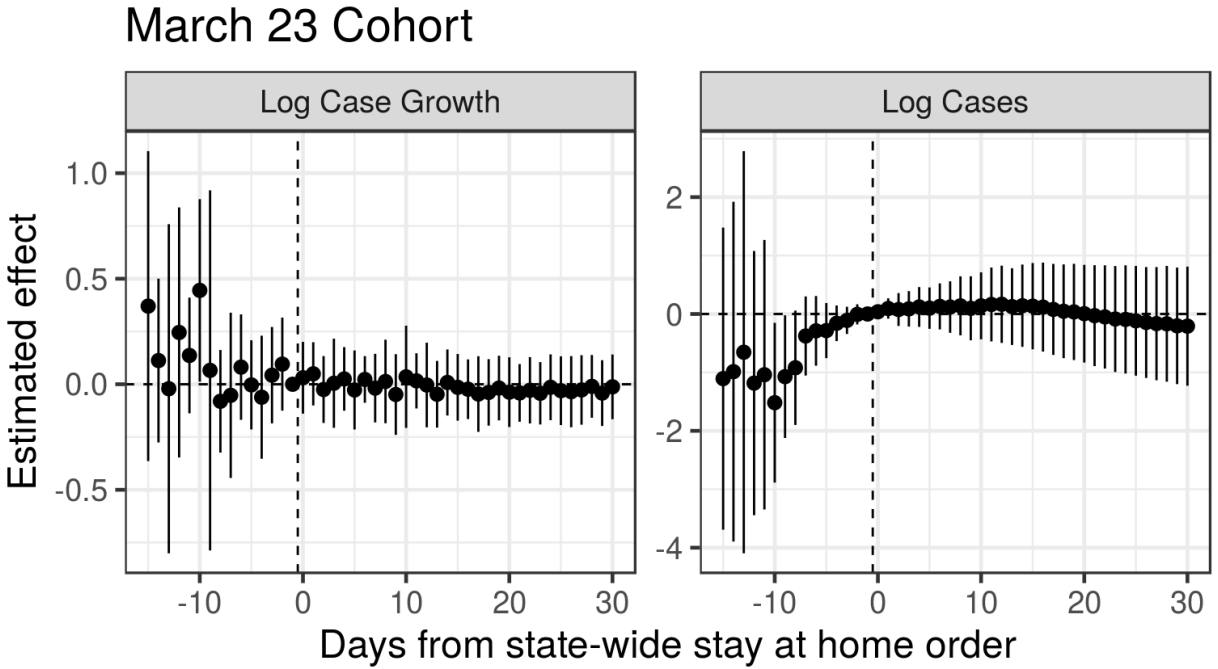

**Figure 2:** Difference-in-differences estimates for the effect of statewide stay-at-home orders on log daily case growth and log cases for states in the March 23 cohort. Standard errors computed via jackknife.

This procedure has several advantages. First, this provides a diagnostic of the parallel trends assumption, similar in spirit to a balance check. The estimated $\widehat{DID}_{kg}$ for $k < -1$ (to the left of the dotted line) are "placebo" estimates of the impact of treatment $k$ periods *before* treatment is enacted; if the parallel trends assumption holds, these should be close to zero. Although this is not a direct test of the actual assumption (since that involves counterfactual outcomes in the post-policy period) assessing the pre-period trends can be thought of as a proxy for evaluating the assumption. As with all diagnostics, these are not a panacea: there is often limited statistical power to detect differences, and noisy estimates around zero do not absolve researchers from making the case for why the assumptions should hold.[23]

As we see in Figure 2, in the week prior to March 23 the placebo estimates are near zero, but two weeks prior there is higher variance and some evidence that growth rates were systematically higher in the treated cohort than the never treated cohort, relative to March 22. For log cases we see even more stark violations of the parallel trends assumption. Relative to the never-treated states, the March 23 cohort saw a larger increase in the number of cases, possibly evidence of time-varying confounding, although with such few units there is a large amount of uncertainty.

Finally, for $k \geq 0$ we estimate a different treatment effect for each period succeeding treatment, without imposing any assumptions on how we expect the treatment effects to phase in or out. From Figure 2 we see that in this single target trial there is insufficient precision to differentiate the effects from zero, let alone to distinguish a trend.

## Nested target trials

**Selecting units**

We now estimate the overall average impact by repeating the trial emulation approach for all 42 states that eventually adopt a stay-at-home order. As above, the first step is to divide treated states into 17 cohorts based on adoption date. For each cohort, we then emulate a single target trial, selecting the same eight never-treated states as comparisons for *every* target trial. Finally, we aggregate results across these target trials.

These are *nested* target trials in the sense that each target trial can have a different starting point and length of follow up.[24] This approach is sometimes known as *stacking* or *event-by-event analysis* in the econometrics literature.[26] The specific approach we implement here is equivalent to that in Abraham and Sun (2020) and Callaway and Sant'Anna (2019) without any covariates.[9,7,22]

**Treatment effect estimation**

In each single target trial, we estimate a series of two-period difference-in-differences estimates for that cohort as in the previous section. There are many ways to aggregate these estimates across cohorts.[7] Here we aggregate based on days since treatment (sometimes called "event time"):

$$\widehat{DID_k} = \frac{1}{n_1} \sum_{g=1}^{G} n_{1g} \widehat{DID_{kg}}$$

where $G$ is the total number of cohorts, $n_{1g}$ is the number of treated units in cohort $g$, and $n_1$ is the total number of treated states. We refer to these as *nested estimates*.

Figure 3 shows the estimates from this approach, both for the effect on log case growth and on log cases. As with the single target trial, estimates to the right of zero are the treatment effects of interest, and estimates to the left of zero are "placebo estimates" of the impact of the treatment *prior* to the treatment itself. For the left panel (log case growth), the placebo estimates for the ten days before a state enacts a stay at home order are precisely estimated near zero; however, placebo effects prior to this are highly variable and show a downward trend over time. This suggests caution in interpreting the negative estimates to the right of zero. For the right panel (log cases), the placebo estimates are even more starkly different from zero, suggesting that this would not be an appropriate analysis and that the estimated effects are likely merely a reflection of these differential trends.

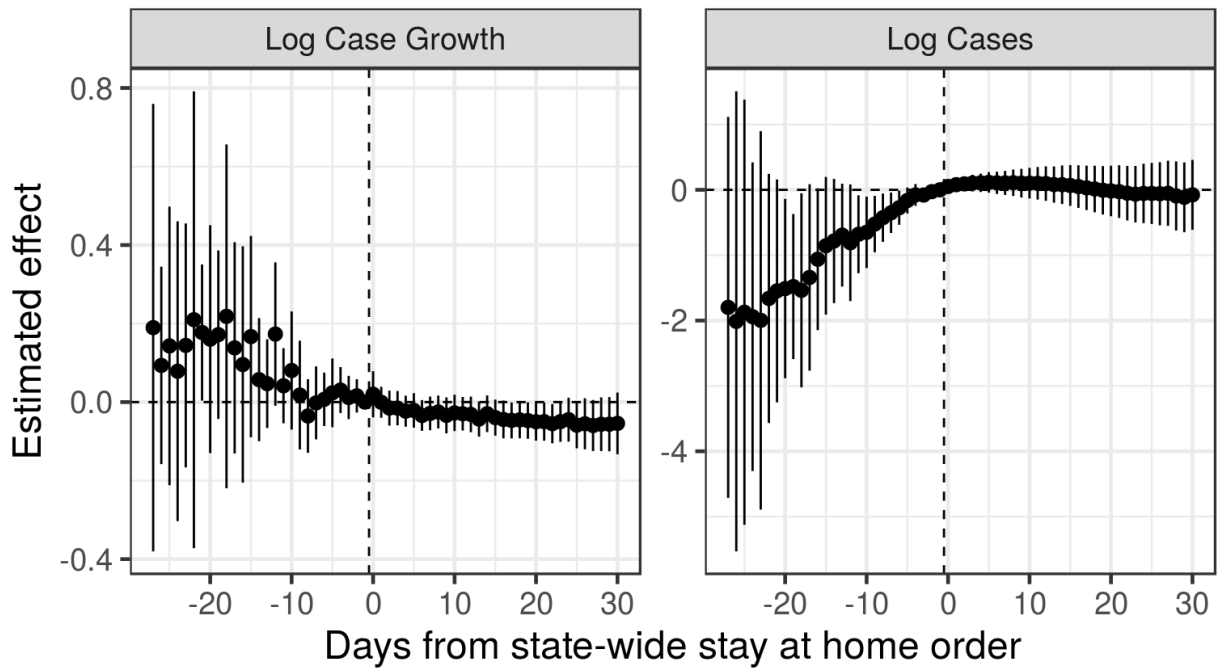

**Figure 3:** Nested estimates of the impact of statewide stay-at-home orders on log cases and log case growth, in calendar time. Standard errors estimated with the jackknife.

## Discussion

Epidemiologists regularly confront settings where multiple jurisdictions adopt a policy over time and the data available are aggregate longitudinal data on those and other jurisdictions. Policy trial emulation provides a principled framework for estimating causal effects in this setting.

The specific approach we advocate is not new; there is a growing literature in statistics and econometrics proposing robust methods for panel data. Here we show that these ideas fit naturally into the trial emulation framework, especially the notion of aggregating across multiple target trials. As a result, we can leverage recent methodological advances to enable more sophisticated estimation that allows for looser assumptions (e.g., parallel trends conditioned on covariates), including inverse propensity score weighting, doubly robust estimation, synthetic controls, and matching. [7] [27] [28] [29] We could also impose stronger modeling assumptions on the time series, e.g., a linear trend, such as in Comparative Interrupted Time Series.[30]

One approach we caution against is the common practice of using regression to fit a longitudinal model to all the data, with fixed effects for state and time. As with individual data with time-varying treatments and confounders, naive regression models can mask important issues. In particular, it has been shown that the coefficient in this pooled model estimates a weighted average over all possible 2x2 difference-in-differences estimates, where the weights can be negative.[22] Moreover, some of these estimates are not in the spirit of trial emulation, e.g., by comparing two states that are entirely post treatment. In practice, these complications mean that the sign of the estimated effect can be flipped relative to the nested estimate. While some approaches, such as "event study models," are less susceptible to this critique,[23] we believe that the trial emulation framework we outline here is more transparent and less prone to error, partly by being explicit about all the causal contrasts.

The issues that we highlight are just some of many major challenges in estimating policy effects more generally, including: differences in the characteristics of states that do and don't implement the policy and challenges in identifying the timing of effects, including limited statistical power.[31] The COVID-19 pandemic adds additional complexities to these policy evaluations.[8] For example, the disease transmission process, and the up to two-week lag in the time from exposure to symptoms, makes it difficult to identify the precise timing of expected effects. Data on outcomes of interest are also limited or problematic; for example, case rates need to be considered within the context of the extent of testing.[32] Finally, methods that do not account for spillovers and contagion are likely to be biased in this setting, and so properly addressing interference is a key methodological and practical concern.

These issues — and the strong underlying assumptions — suggest caution in using difference-in-difference methods for estimating impacts of COVID-19 physical distancing policies. At the same time, the policy trial emulation framework suggests a rubric by which we can assess the quality of evidence presented in these studies. We anticipate that high-quality panel data methods will add to our understanding of these policies, especially when considered alongside other sources of evidence.

**Appendix Figures**

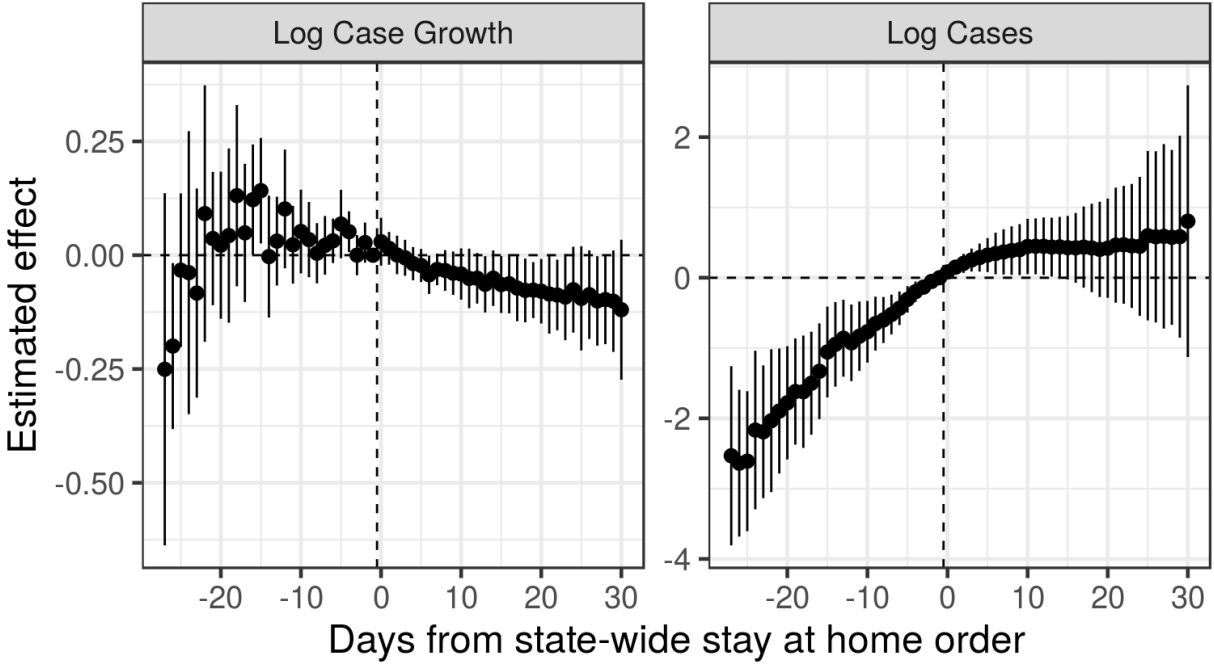

**Appendix Figure 1**: Nested estimates of the impact of statewide stay-at-home orders on log cases and log case growth, in case time. Standard errors estimated with the jackknife.

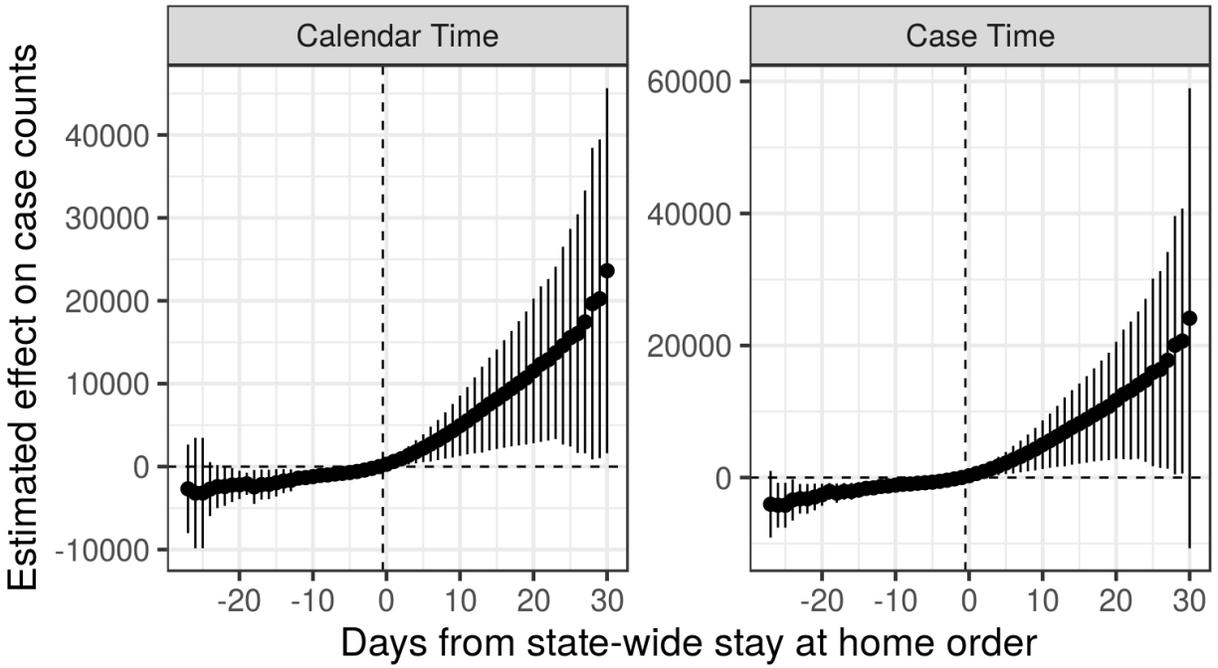

**Appendix Figure 2**: Nested estimates of the impact of statewide stay-at-home orders on raw case counts, in both calendar and case time. Standard errors estimated with the jackknife.

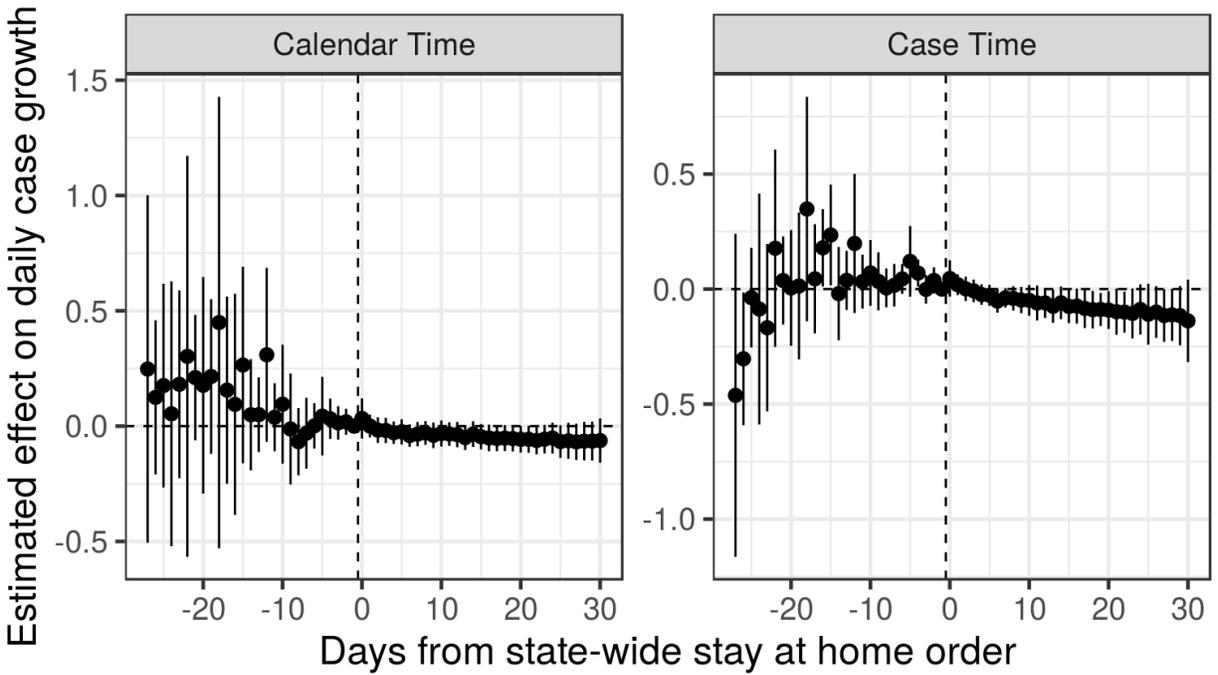

**Appendix Figure 3**: Nested estimates of the impact of statewide stay-at-home orders on raw daily case growth, in both calendar and case time. Standard errors estimated with the jackknife.